%% file: proceedings.tex
\documentclass{sigchi}

\toappear{Permission to make digital or hard copies of all or part of this work for personal or classroom use is      granted without fee provided that copies are not made or distributed for profit or commercial advantage and that copies bear this notice and the full citation on the first page. Copyrights for components of this work owned by others than ACM must be honored. Abstracting with credit is permitted. To copy otherwise, or republish, to post on servers or to redistribute to lists, requires prior specific permission and/or a fee. Request permissions from permissions@acm.org. \\
 {\emph{CHI'14}}, April 26--May 1, 2014, Toronto, Canada. \\
 Copyright \copyright~2014 ACM ISBN/14/04...\$15.00. \\
 DOI string from ACM form confirmation}

\usepackage{balance}  
\usepackage{txfonts}
\usepackage{times}    
\usepackage{microtype}
\usepackage{color}
\usepackage{textcomp}
\usepackage{booktabs}
\usepackage{ccicons}
\usepackage{todonotes}
\usepackage{graphicx}
\usepackage{caption}
\usepackage{subcaption}
\usepackage[ruled,vlined]{algorithm2e}
\usepackage{cite} 
\usepackage{siunitx}
\usepackage{booktabs}
\usepackage{multirow,tabulary}
\usepackage{nth}
\usepackage{adjustbox}

\usepackage{tikz}
\usetikzlibrary{arrows,shapes,trees,fit,positioning,calc}
\usepackage{xspace}

\newcommand{\etal}{\mbox{\em et al.}\xspace}
\newcommand{\ie}{\mbox{\em i.e.}\xspace}

\newcommand{\eg}{\mbox{\em e.g.}\xspace}
\newcommand{\wrt}{\mbox{\em w.r.t.}\xspace}

\newcommand{\etc}{\mbox{\em etc.}\xspace}

\makeatletter
\newcommand*{\rom}[1]{\expandafter\@slowromancap\romannumeral #1@}
\makeatother

\makeatletter
\def\url@leostyle{%
  \@ifundefined{selectfont}{\def\UrlFont{\sf}}{\def\UrlFont{\small\bf\ttfamily}}}
\makeatother
\def\pprw{8.5in}
\def\pprh{11in}

\definecolor{linkColor}{RGB}{6,125,233}
\hypersetup{%
  pdftitle={SIGCHI Conference Proceedings Format},
  pdfauthor={LaTeX},
  pdfkeywords={SIGCHI, proceedings, archival format},
  bookmarksnumbered,
  pdfstartview={FitH},
  colorlinks,
  citecolor=black,
  filecolor=black,
  linkcolor=black,
  urlcolor=linkColor,
  breaklinks=true,
}

\begin{document}

\title{Interactive, Intelligent Tutoring for Auxiliary \\ Constructions in Geometry Proofs} 

\numberofauthors{3}
\author{%
  \alignauthor{Ke Wang\\
    \affaddr{University of California, Davis}\\
    \email{kbwang@ucdavis.edu}}\\
  \alignauthor{Zhendong Su\\
    \affaddr{University of California, Davis}\\
    \email{su@cs.ucdavis.edu}}\\
}

\maketitle

\input{abstract}

\input{intro}

\input{method}

\input{eva}

\input{related}

\input{conclu}

%
%
%
%
%
\balance{}

\bibliographystyle{SIGCHI-Reference-Format}
\bibliography{bibli}

\end{document}

%% file: abstract.tex
\begin{abstract}
Geometry theorem proving forms a major and challenging component in
the K-12 mathematics curriculum. A particular difficult task is to add
auxiliary constructions (\ie, additional lines or points) to aid proof
discovery. Although there exist many intelligent tutoring systems
proposed for geometry proofs, few teach students how to find auxiliary
constructions. And the few exceptions are all limited by their
underlying reasoning processes for supporting auxiliary
constructions. This paper tackles these weaknesses of prior systems by
introducing an interactive geometry tutor, the Advanced Geometry Proof
Tutor (AGPT). It leverages a recent automated geometry prover to
provide combined benefits that any geometry theorem prover or
intelligent tutoring system alone cannot accomplish. In particular, AGPT
not only can automatically process images of geometry problems
directly, but also can interactively train and guide students toward
discovering auxiliary constructions on their own.
We have evaluated AGPT via a pilot study with 78 high school
students. The study results show that, on training students how to
find auxiliary constructions, there is no significant perceived
difference between AGPT and human tutors, and AGPT is significantly
more effective than the state-of-the-art geometry solver that produces
human-readable proofs.

 \end{abstract}

%% file: intro.tex
\section{Introduction}
\label{sec:intro}
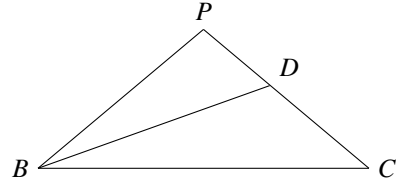
\begin{figure}[!t]
\begin{tikzpicture}[scale=.88]

\coordinate (B) at (2.5,0);
\coordinate (C) at (7.5,0);
\coordinate (P) at (5,2.1);
\coordinate (D) at (6,1.25);

\draw (B) -- (P);
\draw (C) -- (P);
\draw (C) -- (B);
\draw (B) -- (D);

\node [above=.01ex of P] {$P$};
\node [left=.01ex of B] {$B$};
\node [right=.01ex of C] {$C$};
\node [above right=.001cm and .001cm of D] {$D$};

\node[right] at (0,3.5) 
   {
   \begin{minipage}{7.2cm}

   \textbf{Given}: {\footnotesize$BP = CP, \angle{BPC} = 100^\circ, \angle{PBD} = \angle{CBD}$}
   \newline \hspace*{5pt}\textbf{Goal}: {\footnotesize$BC = BD + DP$}
    \end{minipage}
};
\end{tikzpicture}
\vspace*{-3pt}
\caption{Example geometry proof problem.} 
\label{fig:example}
\vspace*{-1pc}
\end{figure}

Geometry is an important, mandatory subject in the secondary school
curriculum. Proof problems form an interesting part of learning
geometry and offer unique challenges to students, both visually and
mathematically~\cite{proofIm1,proofIm2}. They also make geometry one
of the most difficult subjects for students.  The standard format of a
geometry proof problem consists of a geometry figure, a set of
constraints and a goal to be proven. Students are asked to write a
step-by-step deduction using Euclidean
axioms. Figure~\ref{fig:example} depicts a sample problem, which is
used throughout the paper to illustrate our approach. When a problem
requires auxiliary constructions, \ie drawing additional geometrical
elements on the original problem figure, as part of the proof, its
difficulty significantly increases since determining what and where to
draw can be very challenging. No robust and efficient method for
geometry theorem proving with constructions was known~\cite{Matsuda}.

Consequently, easy-to-use and effective intelligent tutoring systems
can help students learn the concepts and practice their problem
solving skills. Unfortunately, the majority of existing geometry proof
tutors~\cite{anderson1985geometry,koedinger1990theoretical,koedinger1993reifying}
do not support teaching and training students how to add
auxiliary constructions. Although the importance of auxiliary
constructions motivates several efforts in the field of artificial
intelligence~\cite{suwa1989acquisition,Matsuda,suautomated}, it is
fair to question their suitability and effectiveness from the
students' learning perspective.  Specifically, even if students have
access to a solution/proof generated by such systems, and may
understand how the auxiliary constructions help solve the given
problem, can they find similar constructions when facing different
problems?  The answer is likely no without the students understanding
the approach the systems adopt to add auxiliary constructions.

In this paper, we propose an interactive geometry tutor, the Advanced
Geometry Proof Tutor (\emph{AGPT}), that targets
auxiliary constructions in geometry proof problems. Our key insight is
that, in order for intelligent tutoring systems to be effective, they
need to utilize the inner workings of powerful solvers (and systems in
general) to train and help users. Specifically, AGPT leverages a
recent automated geometry theorem prover~\cite{suautomated},
\emph{iGeoTutor}, as the core back-end engine to offer the combined
functionalities that neither a geometry theorem prover nor an
intelligent tutoring system alone can accomplish. Traditional
intelligent tutoring systems do not emphasize problem-solving
capabilities and can handle only a limited spectrum of problems, thus
constraining their tutoring support.  In contrast, AGPT can take any
user-supplied problem and leverage the power of the underlying
geometry theorem prover. Compared to automated geometry theorem
provers that focus on automated construction and solution discovery,
AGPT refrains from directly presenting solutions to students. Rather,
it interactively guides students toward discovering a solution on
their own. AGPT borrows the classical pedagogical philosophy from an
intelligent tutoring system's stand point of view --- forcing and
guiding students to think on their own, but providing hints and tips
whenever necessary.

%

We have conducted a pilot study with 78 high school students.  In our
study, we compare AGPT with both human tutors and iGeoTutor, the
state-of-the-art automated geometry theorem prover that produces
human-readable proofs. Our study results show that (1) there is no
significant difference between the tutoring support provided by human
tutors and AGPT, and (2) AGPT is significantly more effective than
iGeoTutor.

%

The rest of this paper is organized as follows. We first describe our
methodology, including a brief summary of the template-based automated
construction search procedure, the technical foundation for this
work. Then, we present our evaluation of AGPT with a pilot
study. Finally, we survey related work and conclude.


%% file: method.tex
\section{Methodology and Design}
\label{sec:method}

\begin{figure}[t!]
\centering
\includegraphics[width=0.48\textwidth]{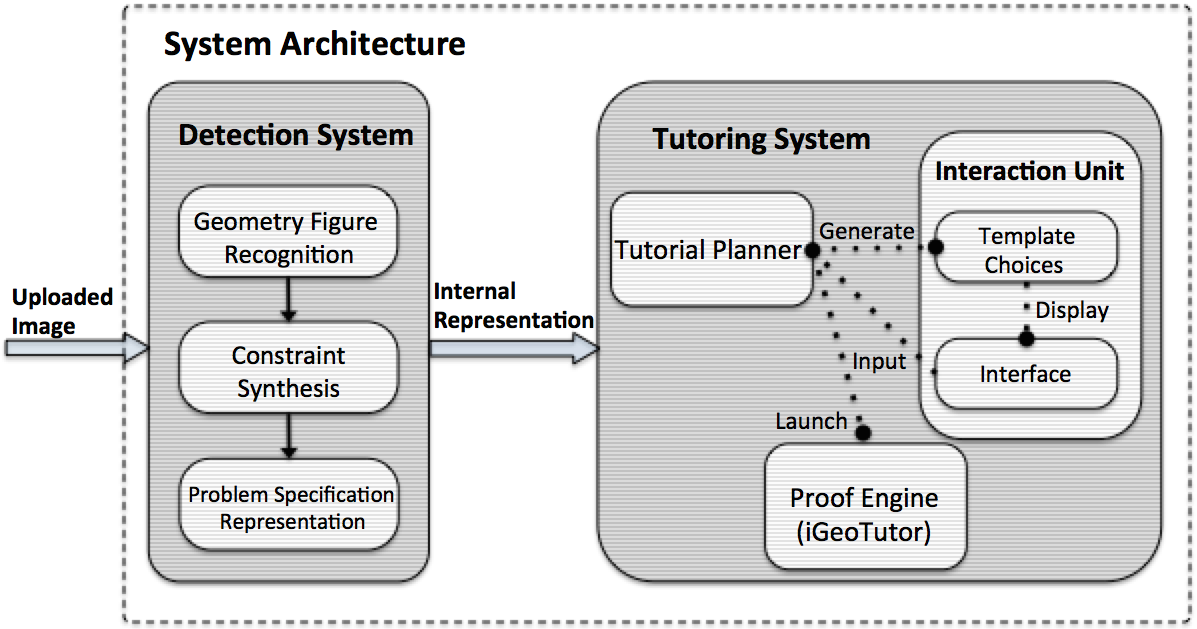}
\caption{System architecture of the geometry tutor.}
\label{fig:arc}
\end{figure}

Figure~\ref{fig:arc} depicts AGPT's system architecture, which
consists of two components: the \emph{detection system} and the
\emph{tutoring system}. In order to help users with their own problems
in the format of image files, AGPT employs the detection system to
parse the geometry figure in the image and then converts it into the
iGeoTutor compliant file format. The tutoring system is responsible
for regulating the interactions between users and the system based on
the discovered constructions and solution by iGeoTutor.


\subsection{Detection System}

As Figure~\ref{fig:arc} shows, the detection system is a three-step
procedure, whose main goal is to relieve the user from the burden of
manually drawing geometry figures. Instead, the user can simply focus
on the tutoring interactions with the system.
  
\subsubsection{Geometry Figure Recognition}

Given a user input image, the first step is to extract the geometry
figure. To this end, we adapt the classical Hough
transform~\cite{HT}. Ideally, we would like the Hough transform to
produce the exact coordinates of points and lines specified in the
uploaded image. However, in practice, those positional data contain
much noise and redundancy. We propose a three-step data filtering
technique, similar to line interpolation, nearby point merging, and
intersection computation, to help accurately recover lines and points
in the original figure.


\begin{enumerate}

\item Given the output from the Hough transform, we first remove
  similar lines by comparing the polar coordinates of each detected
  line, namely one parameter representing the algebraic distance
  between the line and the origin, and another parameter the angle of
  the vector from the origin to this closest point. Care is needed
  when removing the similar lines. According to our experimentation
  with the Hough transform, the detected lines are often shorter than
  the original lines. Thus, instead of removing one or the other when
  two lines are recognized as identical, we form a new line by picking
  the two furthest endpoints among the total four points extracted
  from the two original lines. In other words, we generate a longest
  line with the same polar coordinates by stretching the original
  lines. Finally, we keep the generated line and safely remove any
  duplicates.

\item In terms of the polar coordinates, the data now represent the
  same set of lines as those in the original geometry figure. The
  focus of the second step is to deal with possible discrepancies
  regarding the position of each endpoint. We adopt a simple, direct
  approach by defining the neighborhood of each endpoint and merging
  those that appear in the same neighborhood by calculating the mean.
  
\item The intersection point computation step is to recover
  relationships among points and lines that are not automatically
  inferred by the Hough transform. We consider two types of
  intersections: (1) two lines where one endpoint of one line lies on
  the other line, and (2) two lines crossing each other (\ie in the
  shape of an ``X'').  In the first case, the goal is to decide where
  the target point lies. This can be accomplished simply by computing
  the distance between the target point and a candidate line. However,
  it is worth noting that due to variations on the position of the
  detected point and line by the Hough transform, the point is
  frequently found off the line, such as falling short or exceeding or
  tending toward the ``X'' shape of the intersection. Thus, we introduce a
  threshold to take those minor differences into account. In the
  second case, intersection points can be computed by combining the
  linear functions derived from the existing endpoints and
  incorporating relationships specifying that the intersection falls
  on both lines.
    \end{enumerate}


\subsubsection{Constraint Synthesis}

Apart from recognizing the geometry figure in the user-uploaded image,
the detection system also automatically extracts relevant constraints
for users to select to further increase the user-friendliness of AGPT.

The geometrical elements for which the tutoring system attempts to
infer constraints are segments, angles and special polygons.  A
segment is measured from both the quantitative and positional
perspectives to produce four kinds of predicates, including
perpendicularity, parallelism, equality and addition of length
expressed as $A + B = C$. As for angles, our system only checks the
occurrence of angle bisectors to allow a clean, concise constraint
representation. The angle equality constraints can often be derived
from existing line constraints. The special polygons that we consider
include parallelograms, rectangles, diamonds, squares and equilateral
triangles. To avoid possible duplicate constraints, as soon as a
special polygon is discovered, our system automatically removes the
constraints that can be derived from it.

  \begin{figure}[!t]
\centering
\includegraphics[width=0.47\textwidth]{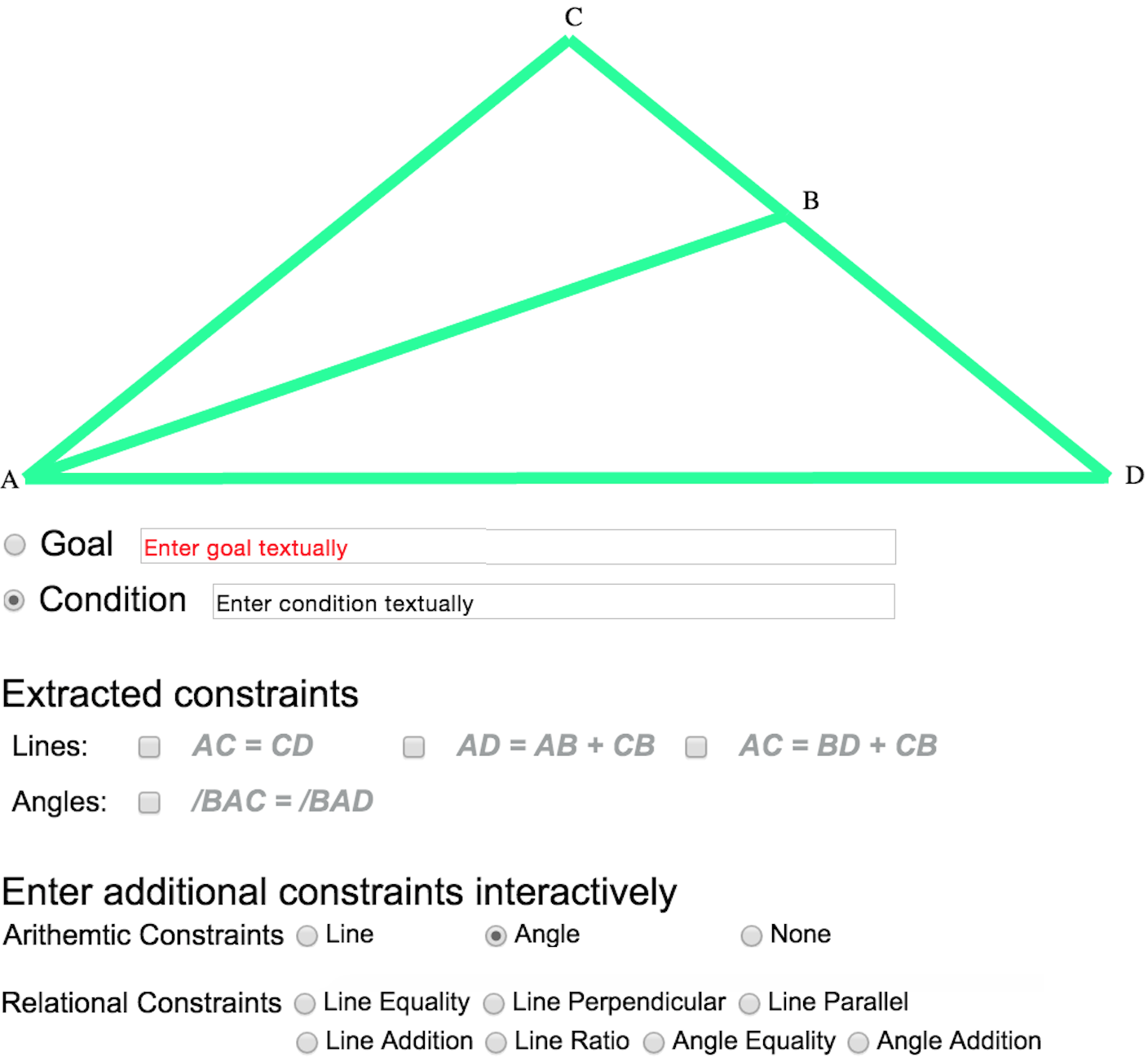}
  \caption{Extracted geometry figure and constraints.}
  \label{fig:inter}
\end{figure}

\subsubsection{Problem Specification Representation}

Given the geometry figure that has been detected and refined along
with its associated constraints on the target elements, our system can
present the problem specification using SVG elements to the user. The
lines displayed on the canvas are all attached with action listeners
to allow the user to interactively supplement missing constraints, if
any. Alternatively, the user may designate a line (or angle) with its
endpoints' identifiers. The user interface depicted in
Figure~\ref{fig:inter} shows the result of the three-step image
processing on a user uploaded picture.

\subsection{Tutoring System}

After the user has specified the problem through the detection system,
the tutoring system will take control and start handling all
interactions with the user. We first briefly review iGeoTutor's
template matching-based proof technique, upon which our tutoring
methodology is built. Then, we present the major decisions that we
have made in designing the tutoring interactions. Next, we describe
the entire interactive tutoring process coordinated by the tutorial
planner.  Finally, we end with a live AGPT's demo session.

\begin{figure}[!t]
\begin{tikzpicture}[scale=.8]

\draw (0,1.3) node[below] {$B$} --
(3,1.3) node[below] {$C$} --
(1.5,4.3) node[above] {$A$} -- cycle;
\draw (1.5,4.3) -- (1.5,1.3) node[below] {$D$};
\draw (1.5,1.5) -- (1.7,1.5) -- (1.7,1.3);
\node[above] at (1,4.7) 
   {
   \begin{minipage}{3.3cm}
     \textbf{Isosceles Triangle (1)}
    \end{minipage}
};

\draw (5,1.3) node[below] {$B$} --
(8,1.3) node[below] {$C$} --
(6.5,4.3) node[above] {$A$} -- cycle;
\node[above] at (7.2,4.7)
{
    \begin{minipage}{3.3cm}
    \textbf{Isosceles Triangle (2)}
    \end{minipage}
};

\draw (0,6.1) node[below] {$C$} --
(2,6.1) node[below] {$D$} --
(1.5,7.6) node[left] {$O$} -- 
(1,9.1) node[above] {$A$} -- 
(3,9.1) node[above] {$B$} -- 
(1.5,7.6) -- 
(1.5,7.6) -- cycle;
\node[above] at (2.2,9.6) 
   {
   \begin{minipage}{5.3cm}
     \textbf{Opposite Triangle}
    \end{minipage}
};
\draw (5,6.1) node[below] {$B$} --
(8,6.1) node[below] {$C$} --
(7.7,9.1) node[above] {$A$} -- cycle;
\draw (6.35,7.6) node[left] {$D$} -- (7.85,7.6) node[right] {$E$};
\node[above] at (7.2,9.6)
{
    \begin{minipage}{3.3cm}
    \textbf{Midpoint Connector}
    \end{minipage}
};

\draw (0,11) node[below] {$B$} --
(1.4,11) node[below] {$C$} --
(0.9,14) node[above] {$A$} --  cycle;
\draw (1.9,11) node[below] {$B'$} -- 
(3.3,11) node[below] {$C'$} -- 
(2.8,14) node[above] {$A'$} -- cycle;
\node[above] at (1.2,14.5)
{
    \begin{minipage}{3.3cm}
    \textbf{Congruent Triangle}
    \end{minipage}
};
\draw (5,11) node[below] {$B$} --
(6.4,11) node[below] {$C$} --
(5.4,14) node[above] {$A$} --  cycle;
\draw (5.4,14) -- (5.4,11) node[below] {$D$};
\draw (5.4,11.2) -- (5.6,11.2) -- (5.6,11);

\draw (6.9,11) node[below] {$B'$} -- 
(8.3,11) node[below] {$C'$} -- 
(7.8,14) node[above] {$A'$} -- cycle;
\draw (7.8,14) -- (7.8,11) node[below] {$D'$};
\draw (7.8,11.2) -- (8,11.2) -- (8,11);
\node[above] at (7.05,14.5)
{
    \begin{minipage}{4.3cm}
    \textbf{Equivalent Area Triangle}
    \end{minipage}
};

\end{tikzpicture}
\caption{The six distilled template figures.} 
\label{fig:TF}
\end{figure}
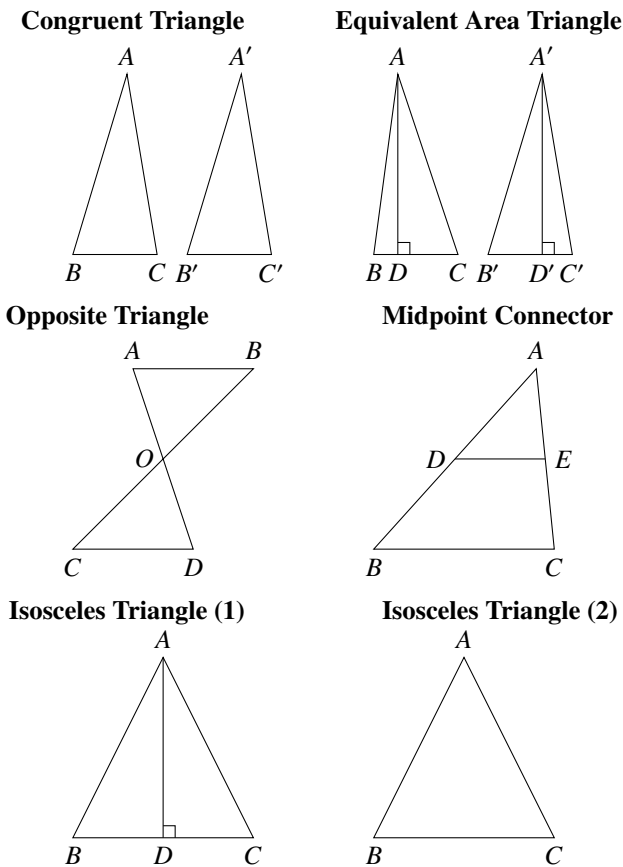

\subsubsection{Background: Template Matching-Based Proof Technique}

iGeoTutor is introduced recently as an automated geometry theorem
prover capable of producing human-readable proofs. It is shown to
significantly outperform the previous state-of-the-art system
\emph{GRAMY}~\cite{Matsuda}. The main technique behind iGeoTutor is
its \emph{template matching-based approach} to tackle the key
challenge in geometry theorem proving, \ie auxiliary construction.
The idea is to guide the search for additional constructions by
\emph{completing} a set of standard shapes, which are called
\emph{templates}. Indeed, the procedure starts by finding suitable
candidate shapes among the templates; Figure~\ref{fig:TF}
depicts the six templates that iGeoTutor currently employs. A shape is a
good candidate if the given problem figure can closely match it in
terms of the given constraints. Once a match is identified, iGeoTutor
can supply the missing constructions to complete the problem
figure. These steps are repeated until a proof is found or a search
limit is reached.

Algorithms~\ref{alg:main} and~\ref{alg:proTem} summarize iGeoTutor's
high-level proof procedure. Function \texttt{Prove} takes as input a
problem figure and infers \texttt{knowledge} that describes the
problem.  If a proof is found without auxiliary constructions, the
function terminates with the proof. Otherwise, the recursive routine
\texttt{ProofByTemplate} is invoked. It first ranks the templates
according to how closely they match the given problem figure. Then, it
iterates over the ranked list of templates, and for each, it adds
auxiliary constructions to complete the current problem figure.  With
the new knowledge in the current problem figure, it attempts to find a
proof again. This recursive process repeats until the system either
finds a proof or reaches a certain search depth. Evaluation results
suggest that a search depth of three is often sufficient in practice.

 \LinesNumberedHidden 
\SetAlFnt{\footnotesize}
\begin{algorithm}[!t]
\DontPrintSemicolon 
\SetKwFunction{proca}{Prove}
\SetKwFunction{ProcTem}{ProofByTemplate}
\SetKwProg{myproc}{procedure}{}{}
\SetKwData{Kn}{knowledge}

\myproc{\proca{GeometryProblem gp}}{ 
	\Begin{

		\Kn $\gets \mathit{KnowledgeExtraction}($gp$)$

		  \If{\Kn.$\mathit{reasoning}()$} {
			    \Return{\Kn.$\mathit{getProof}()$}
  	          }
		  \Else{
     		             \Return{\ProcTem{0, \Kn}}
  		 }
	}
}

\caption{Main proof routine}
\label{alg:main}
\end{algorithm}

\begin{algorithm}[!t]
\DontPrintSemicolon 
\SetKwFunction{proc}{ProofByTemplate}
\SetKwProg{myproc}{function}{}{}
\SetKwData{Kn}{knowledge}
\SetKwData{Dep}{depth}
\SetKwData{Mt}{matchedTemplates}
\SetKwData{OMt}{template}
\SetKwData{Con}{constructions}
\SetKwData{OCon}{construction}
\SetKwData{MaxDepth}{maxDepth}

\myproc{\proc{int \Dep, Knowledge \Kn}}{ 
	\Begin{
		 \If{$\Dep < \MaxDepth$} {
		 	    \Mt $\gets \mathit{TemplateSearchProcedure}(\Kn)$\;

			    \ForEach{\OMt in \Mt}{
			    	\Con $\gets \mathit{SynthesisConstruction}(\OMt)$		    
				
				\ForEach{\OCon in \Con}{
			    		\Kn.$\mathit{addConstruction}$\newline$(\OCon)$\;
				 	\If{\Kn.$\mathit{reasoning}()$} {					
			    			\Return{\Kn.$\mathit{getProof}()$}
			  	         }
			  		\Else{
			   		         \Return{\ProcTem{$\Dep+1$, \Kn}}
			  		 }								
				}
			     }
  	          }		
	}
}

\caption{Proof search via template-based matching}
\label{alg:proTem}
\end{algorithm}

\subsubsection{Design of Interactive Tutoring}

This section presents in detail our design of the tutoring
interactions, including the high-level tutoring strategy, user
interface, feedback mechanism, and the degree to which the tutoring
system manages student interactions.

  
\textbf{\textit{High-Level Tutoring Strategy}}\quad Given the
presented evidence~\cite{suautomated}, iGeoTutor is clearly highly
effective in finding auxiliary constructions in geometry theorem
proving. However, iGeoTutor's underlying technique for 
solving problems is opaque to the students.  Thus, we design tutoring
interactions to expose iGeoTutor's template matching-based approach to
students, and engage them in the problem-solving process so that they
may practice and eventually master the skill.

\begin{figure*}[!ht]
\includegraphics[width=\linewidth]{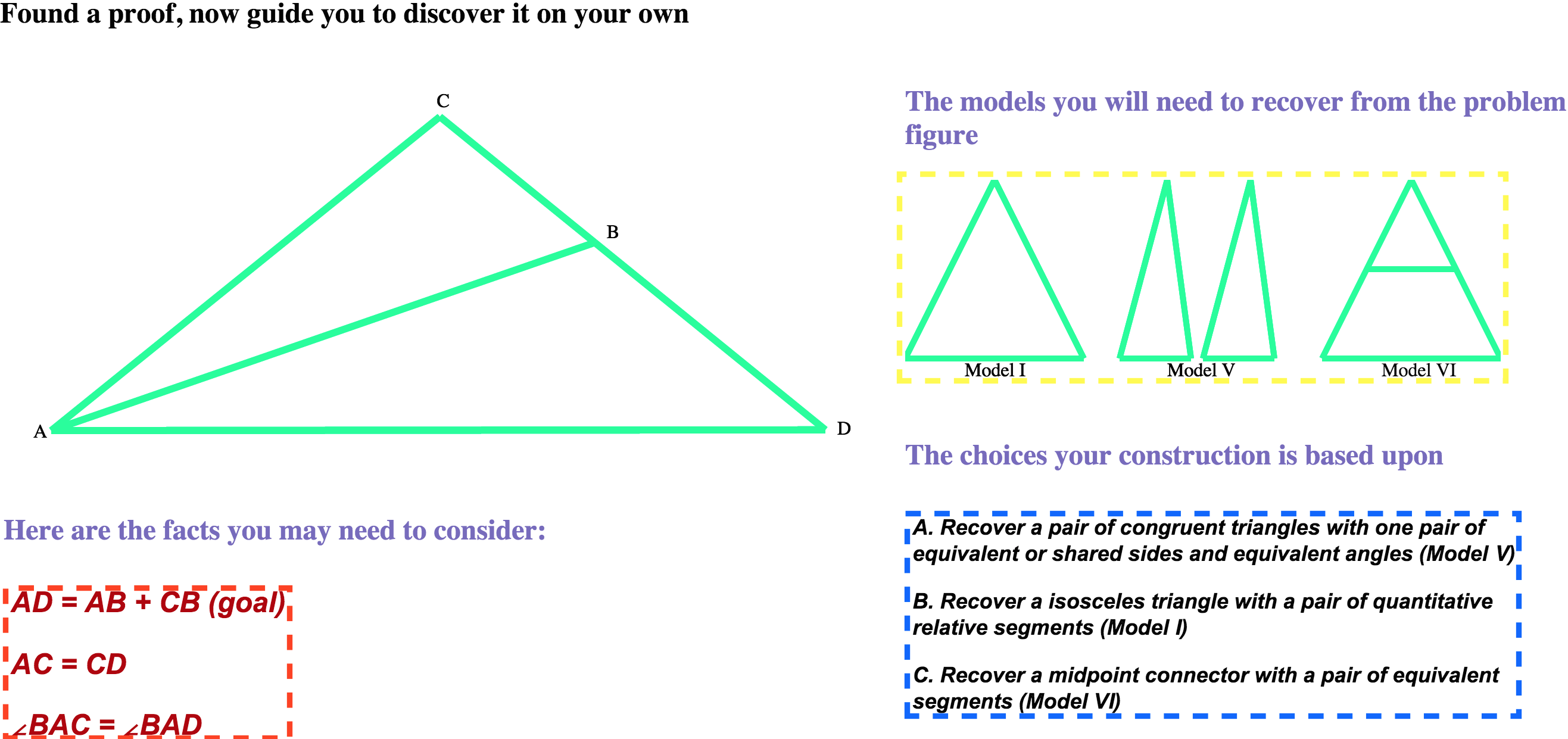}
\caption{A live AGPT tutoring session: template selection.} 
\label{fig:sel}
\end{figure*}

\textbf{\textit{User Interface}}\quad The interactive tutoring
provided by AGPT follows an iterative process. During each iteration
where iGeoTutor automatically finds a template to instantiate, AGPT
pauses and asks the student to manually match a template. Because the
student does not have access to the underlying template figures, AGPT
tailors the exercise of template matching as a two-phase task.  During
the first phase, AGPT presents a multiple choice question by (1)
displaying the exact template figures to be recovered (annotated as
the yellow bounding box in Figure~\ref{fig:sel}) and (2) describing
each choice via a help message on how the template figure can be
recovered (annotated as the blue bounding box). It also reminds the
student the known conditions and conclusion (annotated by the red
bounding box). Figure~\ref{fig:sel} depicts the layout of the entire
interface.

\begin{algorithm}[!t]
\DontPrintSemicolon 
\SetKwFunction{procOne}{InitiateInteraction}
\SetKwFunction{procTwo}{TemplateChoicesGeneration}
\SetKwFunction{procThr}{BackTrackKnowledge}
\SetKwFunction{procFor}{ReasoningWithChoice}
\SetKwProg{myproc}{function}{}{}

\SetKwData{Pr}{proof}
\SetKwData{Kn}{knowledge}
\SetKwData{Tr}{templateTrace}
\SetKwData{Dep}{depth}
\SetKwData{Mt}{matchedTemplates}
\SetKwData{Tp}{turningPoint}
\SetKwData{MaxDepth}{maxDepth}

\SetKwData{OMt}{template}
\SetKwData{Con}{constructions}
\SetKwData{OCon}{construction}

\myproc{\procOne{GeometryProblem gp}}{ 
	\Begin{
	 	\Pr $\gets \proca(gp)$\;
		\Tr $\gets \mathit{GetTemplateTrace}(\Pr)$\;
		\Kn $\gets \mathit{KnowledgeExtraction}(gp)$\;
		\Return{\procTwo{\Kn, \Tr}}
	}	
}
\vspace*{1.15pc}

\myproc{\procTwo{Knowledge \Kn, Templates \Tr}}{ 
	\Begin{
		\Mt $\gets \mathit{TemplateSearchProcedure}(\Kn)$\;
		\Return{$\mathit{SortTemplateChoices}(\Mt, $\newline$\Tr)$}
	}	
}
\vspace*{1.15pc}

\myproc{\procThr{int \Tp, Templates \Tr}}{ 
	\Begin{
		\Kn $\gets \mathit{RestoreKnowledgeBase}(\Tp)$\;
		\Return{\procTwo{Knowledge \Kn, Templates \Tr}}
	}	
}
\vspace*{1.15pc}

\myproc{\procFor{Knowledge \Kn, Templates \Tr, int \Dep, int \Tp, Template choice, Construction model}}{ 
	\Begin{
		 \If{$\neg \mathit{isCorrectConstruction}(\mathit{choice},\mathit{model})$} {
		 	$\Tp \gets \Dep$
		 }
 		
		 $\mathit{SaveKnowledgeBase}(\Kn, \Dep)$
		 \Kn.$\mathit{addConstruction}(\mathit{model})$\;
		 \If{\Kn.$\mathit{reasoning}()$} {
 		  	$\mathit{DisplayCompleteProof}()$
 	       	}
		\ElseIf{$++\Dep == \MaxDepth$}{
			\procThr{\Tp, \Tr}
		}
		\Else{
			$\mathit{choices} \gets \procTwo$\newline$({\Kn, \Tr})$\;
			\If{$\mathit{choices} = \mathit{NULL}$} {
	 		  	\procThr{\Tp, \Tr}
		  	}\;
			\Return{$\mathit{choices}$}
		}
				 
	}	
}

\caption{Interactive tutoring process}
\label{alg:inter}
\end{algorithm}

After the student selects a template, she will be presented with the
second part of the task to specify the constructions to recover her
selected template. In particular, the student needs to make use of the
existing conditions and conclusion (for backward reasoning if
necessary), and obtain the template shape from the original problem
figure by drawing additional lines (\eg, connecting existing points or
introducing points to make new segments). To train students the
concept of template matching, AGPT allows them to draw any segments,
anywhere on the canvas. AGPT also offers support to help students
conveniently convey their intended constructions: (1) allowing
students to draw segments on the problem figure via mouse movements,
(2) automatically adjusting the drawn segments to align with existing
nearby points on the figure, and (3) labeling new intersection points
whenever lines intersect.

The correctness of user input is determined by checking whether the
student's drawing along with some existing segments satisfy the
geometrical constraints associated with the selected template figure.
It is worth mentioning that when AGPT verifies the user construction,
it can tolerate certain degree of inaccuracy in line and point
positioning to increase AGPT's ease-of-use.

Given the construction that the student has specified, if it is not
correct, AGPT will initiate a new iteration of the two-phase
task. Otherwise, AGPT will end the tutoring session.

\textbf{\textit{Feedback Mechanism}}\quad AGPT provides feedback
exclusively in the second phase of the question when students provide
constructions. During the first several failed attempts, AGPT provides
minimal feedback (\eg, ``Try again''). Afterward, if the
student makes further failed attempts, AGPT highlights the relevant
matched segments on the current problem figure.  If the student still
cannot discover a correct construction, AGPT will reveal a solution
and perform the corresponding step.


\textbf{\textit{System Control vs.\ Student Free Exploration}}\quad A
good balance is to let students explore the solution space on their
own, and whenever they need help, AGPT uses the closest solution to
guide students based on their own effort. This approach requires AGPT
to query iGeoTutor for all possible solutions either before the
tutoring session begins or right after each time when students make an
attempt. Both would be computationally expensive and lead to a poor user
experience. To alleviate the computational burden, AGPT requests
iGeoTutor to compute only one solution\footnote{The simplest solution
  is defined by the number of auxiliary lines to be added and templates
  to be matched according to Algorithm~\ref{alg:proTem}.} and use it to
guide students. Note that AGPT does not force students to follow this
solution to maximize their learning experience.
However, AGPT does take special measures to help students stay on the
solution path: (1) it always includes the solution template among the
template choices in the first-phase's multiple choice question, and
(2) it produces hints solely based on the solution construction after
a correct template has been selected and the user seeks feedback from
AGPT.

\subsubsection{Tutorial Planner}
In this section, we describe how the tutorial planner coordinates
different functional modules within the tutoring system to produce
interactions with the student.

\textbf{\textit{Overview}}\quad The tutorial planner begins by
invoking \texttt{Prove} in Algorithm~\ref{alg:main} to find a proof to
the given problem. If a proof is found, it collects the full
construction trace (including template searching and matching) and
uses the trace to direct the subsequent interactions with the student.

Note that the interactions will not be interrupted unless the student
(1) reaches the depth bound without finding a construction, or (2) has
selected a template that leads to a fruitless path (for instance,
there does not exist any matching templates in the next step). In
either case, the tutorial planner will backtrack to the last step
where the student has made a wrong choice or an invalid construction,
and resume from there.

\textbf{\textit{Core Algorithms}}\quad Algorithm~\ref{alg:inter}
describes the main routines that the tutorial planner employs to 
produce the interactive tutoring process.
The function $\mathtt{InitiaiteInteraction}$ starts the interaction
process by invoking $\mathtt{Prove}$ to find the proof's construction
trace. It then invokes the function
$\mathtt{TemplateChoicesGeneration}$ to enumerate the template choices
before presenting them to the student. Once the student has selected a
template, the next step is handled by
$\mathtt{ReasoningWithStudentChoice}$: according to the student's
selection, either another round of template choice generation is
initiated when a solution has not yet been found, or a proof is
synthesized when the student succeeds in finding a correct auxiliary
construction.  Note that $\mathtt{ReasoningWithStudentChoice}$ also
performs backtracking when needed as discussed earlier.

\textbf{\textit{Generating Template Choices}}\quad As shown above, the
function $\mathtt{TemplateChoicesGeneration}$ delegates the task of
finding matched templates to $\mathtt{TemplateSearchProcedure}$. After
all candidate templates have been returned from
$\mathtt{TemplateSearchProcedure}$, up to four templates will be used
and randomly ordered. As mentioned earlier, when the student is still
on the right search path toward a solution,
$\mathtt{TemplateChoicesGeneration}$ needs to include the solution
template from the construction trace.

\subsubsection{An Example Tutoring Session} 
Given the example problem in Figure~\ref{fig:example}, the
student can consult AGPT by first capturing the problem
figure\footnote{Our geometry tutor accepts both printed and
  handwritten figures.} using the camera on their mobile devices, and
then uploading the picture to our system. Next, the student needs to
input the constraints and goal for the given problem figure through
the UI displayed in Figure~\ref{fig:inter}, after which the student
will be presented with the initial round of template selection shown
in Figure~\ref{fig:sel}.

\begin{figure}[!ht]
\includegraphics[width=\linewidth]{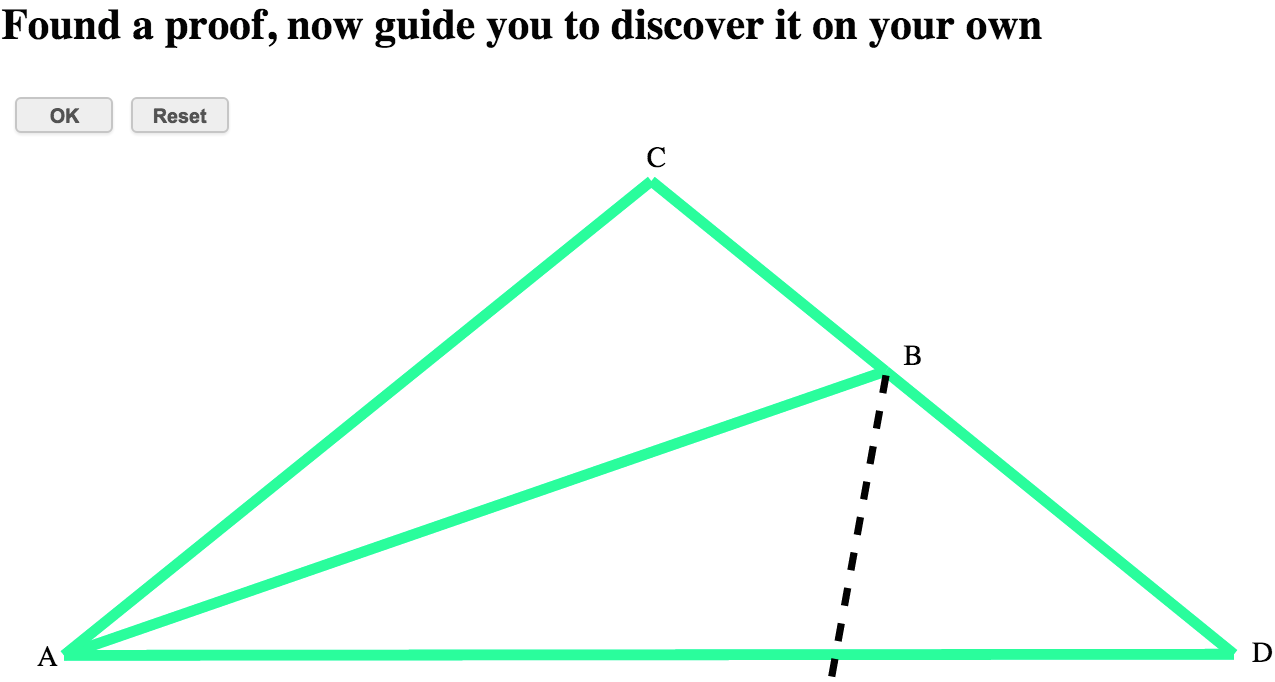}
\caption{A live AGPT's tutoring session: user construction.} 
\label{fig:con}
\end{figure}

As explained earlier, the template selection is presented as a
multiple choice question, placed together with the help message and
template shape. The student, upon selecting a template, will be led to
a canvas shown in Figure~\ref{fig:con} , where she can practice by
drawing freely any auxiliary lines to match the selected template. In
order to differentiate the student's constructions from existing
figure elements, her drawing is displayed as black dashed lines. As
shown in Figure~\ref{fig:con}, the problem figure is automatically
centered on the canvas to best accommodate the student's drawing that
falls outside the problem figure.

AGPT, upon taking the student's constructions into account will either
initiate a new round of template selection based on the student's
added constructions or backtrack to a previous round if a solution is
yet to be found. Otherwise, our system will bring the session to its
end, where the student is offered the opportunity to view the entire
proof. Our submitted supplemental materials contain a demo video to
give a more detailed and concrete illustration.  It is worth to
mention that proof writing on a complete geometry problem figure is
not the focus of our work considering that many of the existing
geometry proof tutors are shown effective both clinically and
empirically. Consequently, we allow students to access the proof
whenever they have added the correct auxiliary constructions.

%% file: eva.tex
\section{Evaluation}

This section presents the pilot study that we conducted to evaluate
the utility of AGPT in helping students learn auxiliary constructions
in geometry proof problems. In particular, we compare AGPT against
human tutors and iGeoTutor, a recent non-interactive geometry theorem
prover~\cite{suautomated}.

 \begin{table*}[!ht]
  \centering
  \begin{tabular}{ c | c | c}
    \hline
    \textbf{Geometry Proof Problem} & \textbf{Auxiliary Construction} & \textbf{Template Matching} \\
    \hline    
    \begin{minipage}{5.6cm} 
    \begin{tikzpicture}[scale=.6]

	\draw (2.9,3) node[above] {$A$} -- (0,.7) node[below] {$B$} -- (1.5,0) node[below] {$E$} -- (2,.7) node[below] {$F$} -- (4,.7) node[below] {$C$} --  cycle;
	\draw (2.9,3) -- (1.5,0);
	\draw (4,.7) -- (2.2,1.45);
	\node[] at (1.9,1.5) {$D$};
	\draw (0,.7) -- (2,.7);
	\draw (2.2,1.45) -- (2,.7);
	


	\node[align=left] at (6.5,2.5) {\small \textit{\underline{Goal:} DF = EF} \\  \small \textit{\underline{Given:} $AE \perp CD$,} \\  \small \textit{$AE \perp BE$,BF = CF}};
	\node[align=left] at (0,4) {\textbf{PRE1.}};		
    \end{tikzpicture}
    \end{minipage}
    
    &
    
    \begin{minipage}{5.6cm} 
    \begin{tikzpicture}[scale=.6]

	\draw (3.9,3) node[above] {$A$} -- (1,.7) node[below] {$B$} -- (2.5,0) node[below] {$E$} -- (3,.7) -- (5,.7) node[below] {$C$} --  cycle;
	\draw (3.9,3) -- (2.5,0);
	\draw (5,.7) -- (3.2,1.45);
	\node[] at (2.9,1.5) {$D$};
	\draw (1,.7) -- (3,.7);
	\draw (3.2,1.45) -- (3,.7);

	\draw[thick,style=dashed] (2.5,0) -- (2.78,-0.125);	
	\draw[thick,style=dashed] (3,.7) -- (2.78,-0.125);		
	\node[] at (3.2,0.4) {$F$};
	\node[] at (3.1,-0.25) {$G$};
	
	\draw[thick,style=dashed] (3.466,1.33) -- (3,.7);		
	\node[] at (3.7,1.6) {$H$};	
	


	\node[align=left] at (7.5,3){\small };
	\node[align=left] at (-1.5,4){\small };		
    \end{tikzpicture}
    \end{minipage}
    
    &

	\begin{tabular}{@{}{l}@{}}
	(\rom{1}). Extend BE and DF to meet at G to \\ 
	to match the template Opposite Trian-\\
	gle with shape BGFDC.\\ 
	\qquad \qquad \qquad \textbf{OR}\\
	(\rom{2}). Extend EF to intersect DC at H to \\ 
	match the template Opposite Triangle \\
	with shape BEFHC.\\ 	
	\end{tabular} 
       \\
       \hline
       
    \begin{minipage}{5.6cm} 
    \begin{tikzpicture}[scale=.6]
        \draw (0,3) node[above] {$A$} -- (0,0) node[below] {$C$} -- (3,0) node[below] {$B$} --  cycle;
        	\draw (0,3) -- (-0.6,1.6) node[left] {$E$};
        	\draw (-0.6,1.6) -- (3,0);
        \node[] at (0.3,1.7) {$D$};	

	\node[align=left] at (4.8,2.4) {\small \textit{\underline{Goal:} $\angle{ABD} = \angle{CBD}$} \\ \small \textit{\underline{Given:} $AC \perp CB$,} \\ \small \textit{AC = CB}, $AE \perp BE$ \\ \small \textit{BD = 2AE}         };
	\node[align=left] at (-1.3,3.88) {\textbf{PRE2.}};
    \end{tikzpicture}
    \end{minipage}
    
    &
    
    \begin{minipage}{5.6cm} 
    \begin{tikzpicture}[scale=.6]
        \draw (0,3) node[above] {$A$} -- (0,0) node[below] {$C$} -- (3,0) node[below] {$B$} --  cycle;
        	\draw (0,3) -- (-0.6,1.6) node[left] {$E$};
        	\draw (-0.6,1.6) -- (3,0);
        \node[] at (0.3,1.7) {$D$};	

        \draw[thick,style=dashed] (-1.3,0)node[below] {$F$} -- (-0.6,1.6);
        \draw[thick,style=dashed] (-1.3,0) -- (0,0);

	\node[align=left] at (4.8,2.4) {\small   };
	\node[align=left] at (-3.6,3.88) {};
    \end{tikzpicture}
    \end{minipage}

    &
    
    	\begin{tabular}{@{}{l}@{}}
	(\rom{1}). Extend AE and BC to meet at F to \\ 
	match the template Isosceles Triangle\\
	(1) with shape BAEF.\\ 
	\end{tabular} 
    
    \\
    \hline
    
    \begin{minipage}{5.6cm} 
    \begin{tikzpicture}[scale=.6]
    \draw (3,4) node[above] {$A$} -- (0,1.5) node[left] {$B$} -- (2.3,1.5) node[below] {$D$} --  (3.55,1.5) node[right] {$C$} -- cycle;
    \draw (3,4) -- (2.3,1.5);


	\node[align=left] at (6.2,3.4) {\small \textit{\underline{Goal:} AB = AC + CD} \\ \small \textit{\underline{Given:} $\angle{C} = 2\angle{B}$,} \\ \small $\angle{BAD} = \angle{CAD}$       };
	\node[align=left] at (0,4.88) {\textbf{PRE3.}};
	\node[align=right] at (0,0) {\textbf{}};
		
    \end{tikzpicture}
    \end{minipage}

    &
        \begin{minipage}{5.6cm} 
    \begin{tikzpicture}[scale=.6]
    
    \draw (3,4) node[above] {$A$} -- (0,1.5) node[left] {$B$} -- (2.3,1.5) node[below] {$D$} --  (3.55,1.5) node[right] {$C$} -- cycle;
    \draw[thick,style=dashed]  (2.3,1.5) -- (3.8,0.3)node[right] {$E$};
    \draw[thick,style=dashed] (3.55,1.5) -- (3.8,0.3);
    \draw (3,4) -- (2.3,1.5);

    \draw[thick,style=dashed] (1.16,2.44)node[left] {$F$} -- (2.3,1.5);

    \node[align=left] at (-2.5,3) {\textbf{}};
	
    \end{tikzpicture}
    \end{minipage}

    &
    
    	\begin{tabular}{@{}{l}@{}}
	(\rom{1}). Introduce a point F on AB to make \\ 
	AF = AC and connect DF to match the \\
	template Congruent Triangle with sha-\\
	pe AFDACD. \\
	\qquad \qquad \qquad \textbf{OR}\\
	(\rom{2}). Extend AC to E such that AE = \\ 
	AB and connect DE to match the tem-\\
	plate Congruent Triangle with shape\\
	ABDAED. 	
	\end{tabular} 
       \\
    \hline

     \begin{minipage}{5.6cm} 
    \begin{tikzpicture}[scale=.6]
    \draw (0,0) node[below] {$B$} -- (3,0) node[below] {$C$} -- (2.5,3.6) node[above] {$D$} --  (1.5,3.6) node[above] {$A$} -- cycle;
    \draw (0,0) -- (2.75,1.8)node[right] {$E$};
    \draw (1.5,3.6) -- (2.75,1.8);

	\node[align=left] at (6.2,3.4) {\small \textit{\underline{Goal:} $\angle{BAE} = \angle{DAE}$} \\ \small $\angle{ABE} = \angle{CBE}$ \\ \small \textit{\underline{Given:} AB = AD + BC,} \\ \small \textit{DE = EC}, $AD \parallel BC$};
	\node[align=left] at (0,4.88) {\textbf{POST1.}};
		
    \end{tikzpicture}
    \end{minipage}

    &

     \begin{minipage}{5.6cm} 
    \begin{tikzpicture}[scale=.6]
    \draw (0,0) node[below] {$B$} -- (3,0) node[below] {$C$} -- (2.5,3.6) node[above] {$D$} --  (1.5,3.6) node[above] {$A$} -- cycle;
    \draw (0,0) -- (2.75,1.8)node[right] {$E$};
    \draw (1.5,3.6) -- (2.75,1.8);
    
    \draw[thick,style=dashed]  (4,0) node[below] {$F$} -- (3,0);
    \draw[thick,style=dashed]  (4,0) --  (2.75,1.8);
    
    \draw[thick,style=dashed]  (5.5,3.6)node[above] {$G$} --  (2.5,3.6);
    \draw[thick,style=dashed]  (5.5,3.6) --  (2.75,1.8);

    \node[align=left] at (-1.5,4.88) {};
		
    \end{tikzpicture}
    \end{minipage}

    &
    	\begin{tabular}{@{}{l}@{}}
	(\rom{1}). Extend AD and BE to meet at G to\\
	match the template Isosceles Triangle\\
	(1) with shape ABEG.\\ 
	\qquad \qquad \qquad \textbf{OR}\\
	(\rom{2}). Extend AE and BC to meet at F to \\
	match the template Isosceles Triangle\\
	(1) with shape BAEF.\\ 	
	\end{tabular} 
       \\
       \hline

     \begin{minipage}{5.6cm} 
    \begin{tikzpicture}[scale=.6]
    \draw (0,0) node[below] {$B$} -- (4,0) node[below] {$C$} -- (2.2,3.1) node[above] {$A$} -- cycle;
    \draw (1.12,1.62)node[left] {$Q$} -- (4,0);
    \draw (3.1,1.55)node[right] {$P$} -- (0,0);
    
    \draw (3,1.48) node[below] {$H$} -- (2.2,3.1);    
    \draw (1.28,1.48) node[below] {$K$} -- (2.2,3.1);    
    
    \draw (1.28,1.48) -- (3,1.48);
    

	\node[align=left] at (5.8,3.4) {\small \textit{\underline{Goal:} $KH \parallel BC$} \\ \small \textit{\underline{Given:} $\angle{ACQ} = \angle{BCQ}$,} \\ \small $\angle{ABP} = \angle{CBP}$, \small $AK \perp CQ$,\\ \small $AH \perp BP$};
	\node[align=left] at (0,4.88) {\textbf{POST2.}};
		
    \end{tikzpicture}
    \end{minipage}
    
    &
         \begin{minipage}{5.6cm} 
        \begin{tikzpicture}[scale=.6]
    \draw (0,0) node[below] {$B$} -- (4,0) node[below] {$C$} -- (2.2,3.1) node[above] {$A$} -- cycle;
    \draw (1.12,1.62)node[left] {$Q$} -- (4,0);
    \draw (3.1,1.55)node[right] {$P$} -- (0,0);
    
    \draw (3,1.48) node[below] {$H$} -- (2.2,3.1);    
    \draw (1.28,1.48) node[below] {$K$} -- (2.2,3.1);    
    
    \draw (1.28,1.48) -- (3,1.48);
    
    \draw[thick,style=dashed] (1.28,1.48) -- (.4,0)node[below] {$D$};    
    \draw[thick,style=dashed] (3,1.48) -- (3.7,0);
    \node[] at (3.6,-.4) {$E$};


	\node[align=left] at (-2.5,4.88) {\textbf{}};
		
    \end{tikzpicture}
    \end{minipage}

    &
    	\begin{tabular}{@{}{l}@{}}
	(\rom{1}). Extend AK to intersect BC at D to \\ 
	match the template Isosceles Triangle\\
	with shape CAKD. Extend AH to int-\\
	ersect BC at E to match the template\\
	Isosceles Triangle (1) with shape \\
	BAHE.\\ 	
	\end{tabular} 
    
    \\
    \hline
    
         \begin{minipage}{5.6cm} 
    \begin{tikzpicture}[scale=.6]
    \draw (-.6,0) node[below] {$B$} -- (2.4,0) node[below] {$D$} -- (3.7,1.55) node[right] {$C$} -- (2.4,3) node[above] {$A$} -- cycle;

    \draw (2.4,0) -- (2.4,3);
    \draw (-.6,0) -- (3.7,1.55);

%
%
    
	\node[align=left] at (6.2,3.4) {\small \textit{\underline{Goal:} $DC \perp AC$} \\ \small \textit{\underline{Given:} $\angle{BAD} = \angle{CAD}$,} \\ \small $AB = 2AC$, $AD = BD$};
	\node[align=left] at (0,4.88) {\textbf{POST3.}};
		
    \end{tikzpicture}
    \end{minipage}
    
    &

             \begin{minipage}{5.6cm} 
    \begin{tikzpicture}[scale=.6]
    \draw (-.6,0) node[below] {$B$} -- (2.4,0) node[below] {$D$} -- (3.7,1.55) node[right] {$C$} -- (2.4,3) node[above] {$A$} -- cycle;

    \draw (2.4,0) -- (2.4,3);
    \draw (-.6,0) -- (3.7,1.55);

     \draw[thick,style=dashed]  (5,0)node[below] {$E$} -- (3.7,1.55);
     \draw[thick,style=dashed]  (2.4,0) -- (5,0);
     
     \draw[thick,style=dashed]  (0.9,1.5)node[left]{$F$} -- (2.4,0);
%
%
    
	\node[align=left] at (-2.2,4.88) {\textbf{}};
		
    \end{tikzpicture}
    \end{minipage}

    &
    
        	\begin{tabular}{@{}{l}@{}}
	(\rom{1}). Introduce a point F on AB to make \\ 
	AF = AC and connect DF to match the \\
	template Congruent Triangle with sha-\\
	pe AFDACD. \\
	\qquad \qquad \qquad \textbf{OR}\\
	(\rom{2}). Extend AC to E such that AE =\\ 
	AB and connect DE to match the tem-\\
	plate Congruent Triangle with shape\\
	ABDAED. 	
	\end{tabular}

    \\
    \hline
  \end{tabular}
  \caption{All the problems used in our evaluation. \textbf{PRE1}-\textbf{PRE3} are the problems on Pre-Test, similarly \textbf{POST1}-\textbf{POST3} are the problems on Post-Test. For problems on the Pre-Test, (I) is AGPT's adopted solution while (II) is discovered by students. The solutions to the Post-Test problems are all discovered by students.}\label{Table:problems}
\end{table*}

\subsection{Participants}

We have contacted the geometry teacher in a neighboring local high 
school, who helped us recruit all ninth grade\footnote{Ninth grade is the 
typical grade level for meeting the Geometry standards in the United 
States.} students (78 in total) for this study. Each participant is given a 
booklet that contains all the geometry theorems that they should find 
helpful during the study.

\subsection{Procedure}

First, all participants took a pre-test. Then, we asked the geometry
teacher to divide the students into three groups (26 students
each). The first group interacted with a human tutor (the same
geometry teacher with more than five years experience in teaching
geometry) regarding the questions on the pre-test (A Group). The
second group used iGeoTutor to explore the solutions to the pre-test
problems (B Group). The third group interacted with AGPT (C Group). We
allocated up to 15 minutes for students to practice and familiarize
with the UIs of iGeoTutor and AGPT (both running on Samsung Chromebook
Plus provided by the local high school), during which students are
allowed to seek help from the study assistants. However, we restrict
the questions to be only UI-related. Students could only work
individually during the tutoring sessions with iGeoTutor or
AGPT. Finally, all students took a post-test. Both the pre-test and
post-test were paper based and contained three geometry proof problems
shown in Table~\ref{Table:problems}.  Each participant was given 30
minutes to complete the pre-test and post-test, and one hour for
interacting with the human tutor (the geometry teacher), iGeoTutor, or
AGPT (excluding the 15 minutes training on the UIs).

\newcolumntype{Y}{>{\centering\arraybackslash}J}
\renewcommand{\arraystretch}{1}

\begin{table*}[!ht]
\begin{center}

\begin{adjustbox}{max width=\textwidth}
\begin{tabulary}{1.5\textwidth}{Y|YYYYYYY|YYYYYYY|YYYYYYY}

\noalign{\hrule height 1pt}
\multirow{2}{*}{\textbf{Student}} 
  &\multicolumn{7}{c|}{\textbf{PRE1}} &\multicolumn{7}{c|}{\textbf{PRE2}} &\multicolumn{7}{c}{\textbf{PRE3}}\\
\cline{2-22}
            &\nth{1} &\nth{2} &\nth{3} &\nth{4} &\nth{5} &\nth{6} &\nth{7} 
            &\nth{1} &\nth{2} &\nth{3} &\nth{4} &\nth{5} &\nth{6} &\nth{7} 
            &\nth{1} &\nth{2} &\nth{3} &\nth{4} &\nth{5} &\nth{6} &\nth{7} \\
\hline
$S001$ &\textit{W} &\textit{W} &\textit{W (B)} &\textit{W (D)} & \textit{M} & \textit{--} & \textit{--} 
             &\textit{W} &\textit{W (D)} &\textit{W (D)} &\textit{I} &\textit{W (B)} & \textit{W} & \textit{C}
             &\textit{C} &\textit{--} &\textit{--} &\textit{--} & \textit{--} & \textit{--} & \textit{--} \\ 

$S002$ &\textit{W} &\textit{W} &\textit{W (B)} &\textit{W} &\textit{W}  &\textit{W (B)}& \textit{W}  
             &\textit{I} &\textit{W} &\textit{W (B)} &\textit{I} & \textit{W} &\textit{W (B)} & \textit{M}
             &\textit{W} &\textit{W} &\textit{W (B)} &\textit{I} & \textit{--} & \textit{--} & \textit{--} \\ 

$S003$ &\textit{W} &\textit{W} &\textit{W (B)} &\textit{I} & \textit{W( D)} & \textit{I} & \textit{--} 
             &\textit{W} &\textit{W} &\textit{W (B)} &\textit{W} & \textit{W} & \textit{W (B)} & \textit{W}
             &\textit{W} &\textit{W (D)} &\textit{W} &\textit{W (B)} & \textit{W} &\textit{I} & \textit{--} \\ 

$S004$ &\textit{W} &\textit{W (D)} &\textit{W} &\textit{W (D)} & \textit{M} & \textit{--} & \textit{--} 
             &\textit{W} &\textit{W} &\textit{W (B)} &\textit{I} & \textit{W} & \textit{W (B)} & \textit{C}
             &\textit{W} &\textit{W} &\textit{W (B)} &\textit{W} & \textit{W (D)} & \textit{M} & \textit{--} \\ 

$S005$ &\textit{I} &\textit{W (D)} &\textit{C} &\textit{--} & \textit{--} & \textit{--} & \textit{--} 
             &\textit{W} &\textit{W} &\textit{W (B)} &\textit{W} & \textit{W (D)} & \textit{M} & \textit{--}
             &\textit{W} &\textit{W (D)} &\textit{I} &\textit{--} & \textit{--} & \textit{--} & \textit{--} \\ 

$S006$ &\textit{W} &\textit{W} &\textit{W (B)} &\textit{W} & \textit{W} & \textit{W (B)} & \textit{W} 
             &\textit{W} &\textit{W} &\textit{W (B)} &\textit{W} & \textit{W} & \textit{W (B)} & \textit{W}
             &\textit{W} &\textit{W} &\textit{W (B)} &\textit{W} & \textit{W} & \textit{W (B)} & \textit{W} \\

$S007$ &\textit{W} &\textit{W (D)} &\textit{W} &\textit{W (B)} & \textit{W} & \textit{C} & \textit{--} 
             &\textit{W} &\textit{W (D)} &\textit{M} &\textit{--} & \textit{--} & \textit{--} & \textit{--}
             &\textit{W} &\textit{W (D)} &\textit{W} &\textit{W (B)} & \textit{W} & \textit{W (D)} & \textit{W} \\ 

$S008$ &\textit{W} &\textit{W} &\textit{W (B)} &\textit{W} & \textit{W} & \textit{W (D)} & \textit{W (B)} 
             &\textit{W} &\textit{W} &\textit{W (B)} &\textit{W} & \textit{W} & \textit{W (B)} & \textit{W} 
             &\textit{W} &\textit{W} &\textit{W (B)} &\textit{W (D)} & \textit{W} & \textit{W} & \textit{W (B)} \\ 

$S009$ &\textit{W} &\textit{W} &\textit{W (B)} &\textit{W (D)} & \textit{M} & \textit{--} & \textit{--} 
             &\textit{W} &\textit{W} &\textit{W (B)} &\textit{W} & \textit{W} & \textit{W (B)} & \textit{W}
             &\textit{W (D)} &\textit{M} &\textit{--} &\textit{--} & \textit{--} & \textit{--} & \textit{--} \\ 

$S010$ &\textit{W} &\textit{W} &\textit{W (B)} &\textit{M} & \textit{--} & \textit{--} & \textit{--} 
             &\textit{I} &\textit{W} &\textit{W (B)} &\textit{W} & \textit{W} & \textit{W (B)} & \textit{W}
             &\textit{W} &\textit{W} &\textit{W (B)} &\textit{W} & \textit{W} & \textit{W (D)} & \textit{W (B)} \\ 

$S011$ &\textit{I} &\textit{W} &\textit{W (B)} &\textit{W} & \textit{W (D)} & \textit{I} & \textit{--} 
             &\textit{W} &\textit{W} &\textit{W (B)} &\textit{I} & \textit{W (D)} & \textit{M} & \textit{--}
             &\textit{I} &\textit{--} &\textit{--} &\textit{--} & \textit{--} & \textit{--} & \textit{--} \\ 

$S012$ &\textit{W} &\textit{W} &\textit{W (B)} &\textit{W} & \textit{W} & \textit{W (B)} & \textit{W} 
             &\textit{W} &\textit{W} &\textit{W (B)} &\textit{W} & \textit{W} & \textit{W (B)} & \textit{W}
             &\textit{W} &\textit{W} &\textit{W (B)} &\textit{C} & \textit{--} & \textit{--} & \textit{--} \\ 

$S013$ &\textit{W} &\textit{W} &\textit{W (B)} &\textit{W (D)} & \textit{W} & \textit{W} & \textit{W (B)} 
             &\textit{I} &\textit{W} &\textit{W (B)} &\textit{C} & \textit{--} & \textit{--} & \textit{--}
             &\textit{I} &\textit{W (D)} &\textit{I} &\textit{--} & \textit{--} & \textit{--} & \textit{--} \\ 

$S014$ &\textit{W} &\textit{W} &\textit{W (B)} &\textit{I} & \textit{W} & \textit{W (B)} & \textit{W} 
             &\textit{W} &\textit{W} &\textit{W (B)} &\textit{I} & \textit{W} & \textit{W (B)} & \textit{C}
             &\textit{W} &\textit{W} &\textit{W (B)} &\textit{W (D)} & \textit{W (D)} & \textit{W} & \textit{W} \\ 

$S015$ &\textit{W} &\textit{W (D)} &\textit{W} &\textit{W (B)} & \textit{W} & \textit{M} & \textit{--} 
             &\textit{W (D)} &\textit{W} &\textit{W} &\textit{W (B)} & \textit{W} & \textit{W} & \textit{W (B)}
             &\textit{W} &\textit{W} &\textit{W (B)} &\textit{W (D)} & \textit{ M} & \textit{--} & \textit{--} \\ 

$S016$ &\textit{M} &\textit{--} &\textit{--} &\textit{--} & \textit{--} & \textit{--} & \textit{--} 
             &\textit{W} &\textit{W} &\textit{W (B)} &\textit{W (D)} & \textit{C} & \textit{--} & \textit{--}
             &\textit{W} &\textit{W} &\textit{W (B)} &\textit{W (D)} & \textit{W} & \textit{W (D)} & \textit{W} \\ 

$S017$ &\textit{W} &\textit{W} &\textit{W (B)} &\textit{W} & \textit{W} & \textit{W (B)} & \textit{W} 
             &\textit{W} &\textit{W (D)} &\textit{W} &\textit{W (B)} & \textit{W} & \textit{W} & \textit{W (B)}
             &\textit{W} &\textit{W} &\textit{W (B)} &\textit{W} & \textit{W (D)} & \textit{W} & \textit{W (B)} \\ 

$S018$ &\textit{W} &\textit{W} &\textit{W (B)} &\textit{W} & \textit{W} & \textit{W (B)} & \textit{W} 
             &\textit{I} &\textit{W} &\textit{W (B)} &\textit{I} & \textit{W (D)} & \textit{W} & \textit{W (B)}
             &\textit{W} &\textit{W (D)} &\textit{I} &\textit{--} & \textit{--} & \textit{--} & \textit{--} \\ 

$S019$ &\textit{I} &\textit{--} &\textit{--} &\textit{--} & \textit{--} & \textit{--} & \textit{--} 
             &\textit{W} &\textit{W} &\textit{W (B)} &\textit{W (D)} & \textit{M} & \textit{--} & \textit{--}
             &\textit{W} &\textit{W (D)} &\textit{W} &\textit{W (B)} & \textit{W (D)} & \textit{W} & \textit{W} \\ 

$S020$ &\textit{W} &\textit{W} &\textit{W (B)} &\textit{W} & \textit{W (D)} & \textit{W (D)} & \textit{M} 
             &\textit{W} &\textit{W} &\textit{W (B)} &\textit{W} & \textit{W (D)} & \textit{W} & \textit{W (B)}
             &\textit{W} &\textit{W} &\textit{W (B)} &\textit{I} & \textit{W} & \textit{W (B)} & \textit{W} \\ 

$S021$ &\textit{I} &\textit{W} &\textit{W (B)} &\textit{W} & \textit{W} & \textit{W (B)} & \textit{W} 
             &\textit{W} &\textit{W} &\textit{W (B)} &\textit{W} & \textit{W} & \textit{W (B)} & \textit{W}
             &\textit{W} &\textit{W} &\textit{W (B)} &\textit{W (D)} & \textit{W} & \textit{W} & \textit{W (B)} \\ 

$S022$ &\textit{W} &\textit{W} &\textit{W (B)} &\textit{W} & \textit{W} & \textit{W (B)} & \textit{C} 
             &\textit{W} &\textit{W} &\textit{W (B)} &\textit{I} & \textit{W} & \textit{W (B)} & \textit{W}
             &\textit{W} &\textit{W} &\textit{W (B)} &\textit{W} & \textit{W} & \textit{W (B)} & \textit{W} \\ 

$S023$ &\textit{W} &\textit{W (D)} &\textit{M} &\textit{--} & \textit{--} & \textit{--} & \textit{--} 
             &\textit{W} &\textit{W} &\textit{W (B)} &\textit{W} & \textit{W} & \textit{W (B)} & \textit{W}
             &\textit{I} &\textit{W} &\textit{W (B)} &\textit{W (D)} & \textit{I} & \textit{--} & \textit{--} \\ 

$S024$ &\textit{W} &\textit{W} &\textit{W (B)} &\textit{I} & \textit{--} & \textit{--} & \textit{--} 
             &\textit{W (D)} &\textit{W} &\textit{W} &\textit{W (B)} & \textit{W} & \textit{M} & \textit{--}
             &\textit{W} &\textit{W} &\textit{W (B)} &\textit{W} & \textit{W} & \textit{W (B)} & \textit{W} \\ 

$S025$ &\textit{W} &\textit{W} &\textit{W (B)} &\textit{W} & \textit{W} & \textit{W (B)} & \textit{W} 
             &\textit{I} &\textit{W} &\textit{W (B)} &\textit{W} & \textit{W} & \textit{W (B)} & \textit{W}
             &\textit{W} &\textit{W} &\textit{W (B)} &\textit{M} & \textit{--} & \textit{--} & \textit{--} \\ 

$S026$ &\textit{W} &\textit{W} &\textit{W (B)} &\textit{W} & \textit{W (D)} & \textit{W} & \textit{W} 
             &\textit{W} &\textit{W (D)} &\textit{I} &\textit{W (B)} & \textit{W} & \textit{W} & \textit{W (B)}
             &\textit{W} &\textit{W} &\textit{W (B)} &\textit{W (D)} & \textit{W (D)} & \textit{W} & \textit{W} \\

\noalign{\hrule height 1pt}
\end{tabulary}
\end{adjustbox}

\caption{Student interaction history.}
\label{table:his}
\end{center}
\end{table*}

\subsection{Student Interactions}
Student in Group A went through a standard class, during which the
geometry teacher explained the pre-test problems and solutions, and
answered their questions. For Group B students, there was little
interaction to be recorded. Students first drew the problem figure and
entered the geometrical constraints, and then waited for a solution,
which they spent the majority of their time studying. We logged
the detailed interaction history for each student using AGPT and
report them in Table~\ref{table:his}.  For all Group C students, none
managed to find a solution to any problem after seven attempts. In
each attempt, facing the two-phase task, a student can have the
following responses:

\begin{itemize}
   \item \textbf{\textit{W}}: The student does not choose the same
     template as the solution template at the same iteration. For
     example, assume that a solution template trace is only an
     instance of Isosceles Triangle (1). If the student selects the
     Opposite Triangle template in her first attempt or Isosceles
     Triangle (1) in her second attempt, in either case the
     student's response will be considered \textit{W}.
   
   \item \textbf{\textit{M}}: The student selects the solution
     template at the right step and obtains a solution construction
     through AGPT's feedback mechanism.
   
   \item \textbf{\textit{C}}: The student selects the solution
     template at the right step and supplies a solution construction
     on his own.
   
   \item \textbf{\textit{I}}: The student selects the solution
     template at the right step and supplies a non-solution
     construction on his own.
\end{itemize}

It is worth to mention that a non-solution template/construction does
not necessarily indicate a wrong solution. In fact, students managed
to find correct alternatives to PRE1 and PRE3, marked as (II) in the
last column in Table~\ref{table:his}.  There are two other symbols
expressed in parenthesis \textit{B}, meaning that AGPT backtracked the
student to where she was diverted away from the solution path.
\textit{D} means that the student pressed the back button to return to
the previous stage.  From Table~\ref{table:his}, we highlight below a
few important observations:

\newcolumntype{Y}{>{\centering\arraybackslash}J}
\renewcommand{\arraystretch}{1}

\begin{table*}[!ht]
\begin{center}

\begin{tabulary}{1.2\textwidth}{Y|YYY|YYY|YYY|YYY|YYY|YYY}

\noalign{\hrule height 1pt}
\multirow{3}{*}{\textbf{Student}} 
  &\multicolumn{6}{c|}{\textbf{GroupA}} &\multicolumn{6}{c|}{\textbf{GroupB}} &\multicolumn{6}{c}{\textbf{GroupC}}\\
\cline{2-19}
  &\multicolumn{3}{c|}{Pre-Test} &\multicolumn{3}{c|}{Post-Test}&\multicolumn{3}{c|}{Pre-Test} &\multicolumn{3}{c|}{Post-Test}&\multicolumn{3}{c|}{Pre-Test} &\multicolumn{3}{c}{Post-Test} \\
\cline{2-19} 
            &\textit{C} &\textit{S} &\textit{P} &\textit{C} &\textit{S} &\textit{P} &\textit{C} &\textit{S} &\textit{P} &\textit{C} &\textit{S} &\textit{P} &\textit{C} &\textit{S} &\textit{P} &\textit{C} &\textit{S} &\textit{P}   \\
\hline
$S001$ &0 &0 &0 &3 &3 &3 &0 &0 &0 &2 &2 &2 &1 &0.3 &0 &3 &2.3 &2 \\
$S002$ &1 &1 &1 &3 &3 &3 &1 &0.2 &0 &2 &1.2 &1 &0 &0 &0 &2 &2 &2 \\
$S003$ &0 &0 &0 &2 &2 &2 &0 &0 &0 &1 &1 &1 &0 &0 &0 &2 &2 &2 \\
$S004$ &0 &0 &0 &2 &2 &2 &0 &0 &0 &1 &1 &1 &0 &0 &0 &2 &2 &2 \\
$S005$ &0 &0 &0 &1 &1 &1 &0 &0 &0 &1 &1 &1 &0 &0 &0 &3 &3 &3 \\
$S006$ &0 &0 &0 &1 &1 &1 &0 &0 &0 &0 &0 &0 &0 &0 &0 &1 &1 &1 \\
$S007$ &0 &0 &0 &0 &0 &0 &0 &0 &0 &0 &0 &0 &0 &0 &0 &2 &2 &2 \\
$S008$ &0 &0 &0 &0 &0 &0 &0 &0 &0 &0 &0 &0 &0 &0 &0 &1 &0 &0 \\
$S009$ &0 &0 &0 &1 &0 &0 &0 &0 &0 &0 &0 &0 &0 &0 &0 &0 &0.6 &0 \\
$S010$ &0 &0 &0 &1 &0.3 &0 &0 &0 &0 &0 &0 &0 &0 &0 &0 &0 &0 &0 \\
$S011$ &1 &0.5 &0 &2 &1.1 &0 &0 &0 &0 &0 &0 &0 &0 &0 &0 &2 &1.3 &1 \\
$S012$ &0 &0 &0 &1 &0.5 &0 &0 &0 &0 &2 &1.3 &1 &0 &0 &0 &0 &0 &0 \\
$S013$ &0 &0 &0 &0 &0 &0 &0 &0 &0 &0 &0 &0 &0 &0 &0 &1 &0.2 &0 \\
$S014$ &0 &0 &0 &0 &0 &0 &0 &0 &0 &1 &0.2 &0 &0 &0 &0 &1 &0.4 &0 \\
$S015$ &0 &0 &0 &0 &0 &0 &1 &0.4 &0 &2 &0.8 &0 &0 &0 &0 &0 &0 &0 \\
$S016$ &0 &0 &0 &1 &1 &1 &0 &0 &0 &0 &0 &0 &0 &0 &0 &2 &1.5 &1 \\
$S017$ &0 &0 &0 &1 &0.3 &0 &0 &0 &0 &0 &0 &0 &0 &0 &0 &0 &0 &0 \\
$S018$ &0 &0 &0 &2 &1.5 &1 &0 &0 &0 &0 &0 &0 &0 &0 &0 &0 &0 &0 \\
$S019$ &0 &0 &0 &1 &0.3 &0 &0 &0 &0 &1 &0.2 &0 &1 &0.6 &0 &2 &1.1 &0 \\
$S020$ &0 &0 &0 &0 &0 &0 &0 &0 &0 &0 &0 &0 &0 &0 &0 &1 &1 &1 \\
$S021$ &0 &0 &0 &0 &0 &0 &0 &0 &0 &0 &0 &0 &0 &0 &0 &1 &0.6 &0 \\
$S022$ &1 &1 &1 &3 &3 &3 &1 &0.3 &0 &2 &0.6 &0 &0 &0 &0 &0 &0 &0 \\
$S023$ &0 &0 &0 &1 &0.2 &0 &0 &0 &0 &0 &0 &0 &0 &0 &0 &0 &0 &0 \\
$S024$ &0 &0 &0 &0 &0 &0 &0 &0 &0 &0 &0 &0 &0 &0 &0 &2 &2 &2 \\
$S025$ &0 &0 &0 &0 &0 &0 &0 &0 &0 &0 &0 &0 &0 &0 &0 &1 &0.6 &0 \\
$S026$ &0 &0 &0 &0 &0 &0 &0 &0 &0 &0 &0 &0 &0 &0 &0 &0 &0 &0 \\
\noalign{\hrule height 1pt}
\end{tabulary}

\caption{Pilot study results.}
\label{table:res}
\end{center}
\end{table*}

\begin{itemize}
\item There are twenty-one students who solved at least one problem,
  among whom fifteen found solutions through AGPT's feedback
  mechanism, indicating that the feedback was generally helpful. Five
  of the fifteen students did not give a single correct construction
  on their own, and relied completely on AGPT's feedback.  In
  addition, eight students worked out different solutions from AGPT's,
  likely indicating that these students engaged in thoughtful
  problem-solving.

\item There were 452 attempts (392 ended up in \textit{W}) and 153
  redirections (103 backtrackings with 50 self-control) made in total
  by group C students. The data suggests that it takes each student on
  average almost six attempts and two redirections to solve one
  problem. This provides strong evidence that students have
  experienced through a difficult process. More importantly, AGPT's
  support of encouraging students to explore their own ideas has
  benefited them according to their performance on the post-test.

\item Some students displayed noticeable frustrations throughout the
  course of the study. They likely ran out of patience after seven
  attempts and gave up. How to incorporate richer support without
  compromising students' learning would be interesting future work.
\end{itemize}

\subsection{Results}
Table~\ref{table:res} compiles each student's performance on the
Pre-Test and Post-Test. In particular, \textit{C} represents the
number of correct constructions a student found, \textit{S} represents
the number of correct solutions a student was able to work out, and on
top of \textit{S}, \textit{P} introduces a new measure of recording
students' performance which we will explain in detail next.

\subsection{Analysis}

All students did better on the post-test than the pre-test according
to Table~\ref{table:res}.  We have run a paired $t$ test to compare
each participant's performance (in terms of number of solved problems)
across the two test sets within each group. The results show that all
three groups of students' post-test performance is significantly
better than that of the pre-test ([t(25) = 3.2609, p = 0.0032], [t(25)
  = 2.5733, p = 0.0164], and [t(25) = 3.5634, p = 0.0015] for Group A,
B and C respectively).

Using the improvement score, defined as the difference between the
number of correctly solved problems on the post test over the
pre-test, we have run an $\mathit{ANOVA}$ test across the three groups
of students. The results do not reveal any significant difference
among the three interaction approaches ([$F(2, 75) = 2.13, P =
  0.13$]).  Based on an inspection of the post-test student solutions,
we conjecture that the results were due to: (1) the number of problems
on both tests is too few such that the (in)correctness of each problem
would lead to significant differences in the scores, and (2) students
still have difficulties in solving proof problems even on problems
that do not require auxiliary constructions. To mitigate these two
factors, we awarded partial credits for students' incomplete proofs
(\wrt to the \textit{P} column in Table~\ref{table:res}).  Partial
credits were only given to solutions with correctly identified
auxiliary constructions. The rationale is that student's failure in
completely solving a problem with the correct auxiliary constructions
is due to the student's lack of knowledge and skills in proof
construction rather than auxiliary construction. The criterion used to
grade partial solutions is based on comparing students' answers with
the reference solutions. In particular, we calculate the ratio of the
number of correct proof steps students did manage to work out and the
total number of steps.

Taking partial credits into account, we recomputed the pre-test and
post-test scores for each student registered under the \textit{P}
column in Table~\ref{table:res}.  Then, we ran an $\mathit{ANOVA}$ test
on the improvement scores again. The results do reveal significant
difference among the three interaction approaches ([$F(2, 75) = 3.23,
  P = 0.045$]). Post hoc Tukey test indicated no significant
difference between Group A group and Group B, nor between Group A
group and Group C, whereas the difference between Group B and Group C
is significant (P \textless 0.05). To conclude based on the two
$\mathit{ANOVA}$ tests we ran, AGPT is shown to be significantly more
effective than iGeoTutor in helping students learn auxiliary
constructions, even though there is no significant difference detected
between AGPT and the human tutor.

\subsection{Discussion}

The results from our relatively large scale study were consistent with
what we had hypothesized. Below we summarize reasons for AGPT's
effectiveness. First, iGeoTutor's powerful underlying proof strategy
and technique faciliated AGPT to be successful. More importantly,
iGeoTutor alone is insufficient, as evidenced by Group B students in
the study who showed no improvement when they approached new, but
similar problems. Instead, Group C students benefited significantly
more from AGPT, our interactive tutoring system built upon iGeoTutor,
and designed to train students its underlying techniques via template
matching and interactive construction of auxiliary elements. Finally,
AGPT's tutoring style of leading students to go through a challenging,
engaging, and thoughtful process might have frustrated some of them
occasionally, but left a positive impact overall in helping them
learn.

%% file: related.tex
\section{Related work}
\label{sec:re}

This section surveys closely related work on automated geometry
theorem provers and intelligent geometry proof tutors.

\subsection{Automated Geometry Theorem Proving} 
A series of automated geometry proof systems based on decision methods
have dominated the field for decades. Notable work includes Wu-Ritt's
characteristic set method~\cite{ChouSC,ChouSC1,ChouSC2}, the Gr\"obner
method~\cite{Buchberger,Kutzler}, the resultant method~\cite{Kapur},
the elimination method~\cite{Wang}, and the parallel numerical
method~\cite{Zhang}.  The general approach behind these provers is to
use sophisticated algebraic theories to determine the validity of
algebraic formulations of the geometry problems. While they are
powerful in handling a large number of nontrivial theorems in
practice, they have significant drawbacks for use in education. The
most important is that the generated proofs are unnatural and
incomprehensible as these systems do not approach geometry problems
like how secondary school students are taught to.

Recognizing the weaknesses of the aforementioned systems, Chou has
done seminal work on the automated generation of human-understandable
proofs for geometry proof problems. Their work is the pioneer in
attempting to bridge the gap between automated geometry theorem
proving to intelligent geometry proof tutoring. The proposed area
method~\cite{Chou} has been the most powerful geometry theorem proving
algorithm in this domain. However, the work has only limited success
in educational settings because the generated proofs follow
specialized, nonstandard area axioms rather than the standard
Euclidean axioms from traditional geometry textbooks.

GRAMY~\cite{Matsuda} designed by Matsuda \etal is the first proof
system capable of generating human-readable proofs using Euclidean
axioms that students learn in schools. However, as the recent work on
iGeoTutor~\cite{suautomated} shows that GRAMY's approach is quite
inexpressive, and also ineffective and can cause combinatorial
explosion when multiple construction steps are required.

\subsection{Intelligent Geometry Proof Tutor} 

The application of intelligent tutoring systems in the domain of high
school geometry proof problems dates back decade ago.  Among those
geometry proof
tutors~\cite{anderson1985geometry,koedinger1990theoretical,koedinger1993reifying},
Geometry Tutor~\cite{anderson1985geometry} is one of the earliest that
are based on formal geometry rules to solve problems as well as to
teach students. A major weakness of Geometry Tutor (as well as its
companions) is that they cannot handle problems that require auxiliary
constructions; the rules designed in the knowledge are based on the
assumption that a given geometry figure is sufficient for constructing
a proof. Cobo~\etal propose AgentGeom~\cite{cobo2007agentgeom} to
offer students the cognitive and metacognitive support to help them
develop problem solving and mathematical reasoning skills. A major
advantage of AgentGeom over Geometry Tutor is its ability to capture
the students' thought process by adopting discursive and graphic style
of interactions. Additionally, it can feed students hints that match
their state of mind if necessary.  Although AgentGeom's corpus does
contain problems that require auxiliary constructions, its tutoring
activities only follow hard-coded proofs, and more importantly do not
teach students a general, systematic approach for finding auxiliary
constructions. Perhaps the closest to AGPT is Matsuda
\etal~\cite{matsuda1998diagrammatic,Matsuda1}'s systems, which focus
on teaching students how to find auxiliary constructions in geometry
proof problems.
%
%
%
%
%
%
However, both systems' tutoring capabilities are based on GRAMY's
auxiliary construction procedure and thus suffers from poor efficiency
and expressivity as mentioned earlier, especially when dealing with
challenging problems. In addition, neither demonstrates student
learning gains from the interactive tutoring support.  In contrast,
AGPT builds upon the state-of-the-art technology for auxiliary
constructions, and our pilot study shows that AGPT helps students
acquire the skills to find auxiliary constructions as effectively as
human tutors, and significantly more effectively than the geometry
theorem prover alone.
%
%
%
%
%
%
%
%

%% file: conclu.tex
\section{Conclusion}
\label{sec:co}

In this paper, we have presented AGPT, a geometry tutoring system that
leverages its underlying geometry theorem prover to interactively
train and guide students on discovering auxiliary constructions on
their own. AGPT's effectiveness is substantiated by our pilot study
with 78 high school students.  We envision that AGPT's high-level
conceptual approach of disclosing and imparting the inner working of
powerful solvers to students can be fruitfully adapted to other
important educational topics, such as algebra, calculus, physics,
\etc.

%
%
%
%

Our immediate future work is to deploy AGPT to the participating high
schools in our pilot study and continue refining our system based on
student and teacher feedback.

